\documentclass[11pt]{article}

\usepackage[margin=1in]{geometry}
\usepackage{tgtermes}
\usepackage{amsthm}
\usepackage{newtxtext}
\usepackage{newtxmath}
\usepackage{bm}
\usepackage{endnotes}
\usepackage{mathtools}
\usepackage{xcolor}
\usepackage{enumitem}
\usepackage{graphicx}
\usepackage{setspace}

\providecommand{\argmax}{\operatorname*{arg\,max}}

\providecommand{\cvar}{\operatorname{CVaR}}

\newcommand{\E}[1]{\mathbb{E}\!\left[#1\right]}
\newcommand{\pr}[1]{\mathbb{P}\!\left(#1\right)}
\newcommand{\one}[1]{\mathbf{1}\!\left\{#1\right\}}

\newcommand{\paren}[1]{\left(#1\right)}
\newcommand{\brackets}[1]{\left[#1\right]}
\newcommand{\braces}[1]{\left\{#1\right\}}

\newcommand{\R}{\mathbb{R}}

\newcommand{\bbI}{\mathbb{I}}
\newcommand{\bbA}{\mathbb{A}}

\newcommand{\defeq}{\triangleq}

\newenvironment{myproof}{\begin{proof}}{\end{proof}}

\usepackage{natbib}
\bibpunct[, ]{(}{)}{,}{a}{}{,}

\theoremstyle{plain}
\newtheorem{theorem}{Theorem}
\newtheorem{lemma}{Lemma}
\newtheorem{corollary}{Corollary}
\newtheorem{assumption}{Assumption}
\theoremstyle{definition}
\newtheorem{definition}{Definition}
\theoremstyle{remark}
\newtheorem{remark}{Remark}

\title{Risk of Bad Tails: CVaR-Aware Pandora's Box and Prophet Inequalities}
\author{Jingwei Ji\\
Management Science and Engineering\\
Stanford University\\
\texttt{jingwei.ji@stanford.edu}}
\date{}

\begin{document}

\maketitle

\begin{abstract}
We study Conditional Value-at-Risk (CVaR) variants of two canonical sequential decision problems: Pandora's box and the prophet inequality. For Pandora's box, the risk-aware problem retains an elegant Weitzman-style index solution after a one-dimensional variational reduction. For the prophet inequality, the picture is different: for every CVaR level \(\alpha\in(0,1)\), no positive constant approximation guarantee can hold without additional distributional structure, in sharp contrast with the risk-neutral case \(\alpha=1\), and we characterize the tight instance-dependent guarantee. Already in two-item hard instances, the prophet's CVaR benchmark can be made arbitrarily large while every online policy's CVaR remains relatively bounded. This impossibility is due to the nature of CVaR objective: it measures only the worst \(\alpha\)-fraction of outcomes, so any compromise an online policy makes to preserve the chance of a large payoff in the upper \((1-\alpha)\)-fraction might not help its CVaR. It turns out that some additional distributional structure restores a uniform result: under continuous reward distributions satisfying an increasing-failure-rate-average (IFRA) condition, a threshold policy achieves an explicit constant bound.
\end{abstract}


%%%%%%%%%%%%%%%%%%%%%%%%%%%%%%%%%%%%%%%%%%%%%%%%%%%%%%%%%%%%%%%%%%%%%%
\section{Introduction}
\label{sec:intro}

Sequential decision problems have clean risk-neutral answers in several canonical models: in Pandora's box \citep{weitzman1979optimal}, Weitzman's index rule is optimal, and in the classical prophet inequality \citep{krengel1977semiamarts,krengel1978semiamarts}, a simple stopping rule captures a constant fraction of the prophet's expected payoff. These benchmarks are stated in expectation; this paper asks what survives when the criterion is the lower-tail Conditional Value-at-Risk (CVaR) \citep{brown2007large} at level \(\alpha\in(0,1)\), with the risk-neutral case recovered at \(\alpha=1\). 
This question is natural in selection problems where average performance is not enough: a decision maker may want to avoid a choice that looks attractive ex ante but performs poorly in the realized lower tail. In hiring, for example, the concern is not only to find a strong candidate on average, but also to guard against selecting someone who appears promising and later turns out to be a poor fit.
We make several interesting discoveries, which we summarize below. 

For Pandora's box, the risk-aware problem stays exactly tractable: a one-dimensional variational reduction turns it into a family of ordinary Pandora instances, so the optimal policy retains an exact Weitzman-style index solution.

For the prophet inequality, the picture is surprisingly different: for every CVaR level \(\alpha\in(0,1)\), no positive constant approximation guarantee can hold without additional distributional structure, in sharp contrast with the risk-neutral case \(\alpha=1\), where the classical constant-fraction guarantee holds; nevertheless, we manage to characterize a \textit{tight} instance-dependent multiplicative factor guarantee.
The carefully crafted hard instance reveals a fundamental obstruction from the nature of the CVaR objective. 
The construction has only two items. 
Roughly speaking, the first item is designed to have many possible reward values, with rarer values chosen to be proportionally larger, so that adding more possible values increases the prophet's CVaR benchmark.
The second item pays zero with probability \(\alpha\) and a very large value otherwise; hence the prophet's CVaR is governed by the branch in which the second item is zero. 
The online policy must decide whether to take the first item before knowing whether the second item will be zero; if it rejects the first item to preserve the chance of the very large second-item payoff, that upside occurs outside the worst \(\alpha\)-fraction and therefore does not improve its CVaR.
However, there is still a silver lining: increasing-failure-rate-average (IFRA), a classical shape restriction from reliability theory \citep{kochar1987partial,elBarmi2021estimation}, turns out to be sufficient to restore a positive constant multiplicative approximation guarantee.
Intuitively, this additional structure controls the reward tail so that a threshold policy can capture a nontrivial portion of the prophet's CVaR benchmark. 

Technically, the novelty is in how we use the variational representation of CVaR \citep{rockafellar2000optimization,rockafellar2002conditional,pflug2000remarks}. For Pandora's box, fixing the variational scalar does not lead to an augmented dynamic program; instead, it recovers a risk-neutral Pandora instance with capped and rescaled rewards, preserving the index structure up to a one-dimensional outer optimization. For the prophet problem, the same scalar becomes a cap that must be coupled with the stopping threshold, yielding the tight threshold-tail coefficient. 
Moreover, we identify the mechanism through which IFRA certifies a good threshold policy: either the payoff secured at the threshold is already large enough, or the rewards above the threshold still retain enough of the prophet's lower-tail value.

\subsection{Related Work}
\label{sec:related}

\noindent\textbf{Static CVaR.} CVaR is the canonical coherent risk measure in the sense of \cite{artzner1999coherent}; its tractability rests on the variational identity of \cite{rockafellar2000optimization,rockafellar2002conditional} and \cite{pflug2000remarks}, with \cite{shapiroDR2021} as the textbook treatment. The closest dynamic precedent is the risk-sensitive MDP literature: \cite{bauerle2011markov} reduce CVaR-minimization to an MDP on an augmented state space, and \cite{bauerle2014more,bauerle2021spectral,bauerle2024survey} extend and survey this approach. Related algorithmic approaches include \cite{chow2014algorithms,chow2015risk,tamar2015optimizing,tamar2017sequential,haskell2015convex,petrik2012approximate}. For broader stochastic-programming decomposition methods with adaptive sampling, see \cite{zhang2025adaptive}.

\noindent\textbf{Pandora's box variants.}
Recent Pandora extensions include descending-price reformulations \citep{kleinberg2016descending}, combinatorial generalizations \citep{singla2018price}, non-obligatory inspection \citep{doval2018whether,beyhaghi2019pandora,beyhaghi2023pandora,fu2023pandora}, order constraints \citep{boodaghians2023pandora}, correlated prizes \citep{chawla2020pandora,gergatsouli2023weitzman}, contextual settings \citep{atsidakou2024contextual}, and partial inspection \citep{aouad2026pandora}; \cite{beyhaghi2023sigecom} surveys recent developments. Related sequential information-acquisition and selection models also appear in decision-analysis work on technology adoption \citep{ulu2009uncertainty}, Bayesian sequential search with knowledge transfer \citep{huh2025uncertain}, and stochastic hiring pipelines \citep{epstein2024selection}. To our knowledge, no prior paper studies CVaR or other coherent-risk objectives within Weitzman's framework.

\noindent\textbf{Prophet inequality variants.}
Beyond the foundational \cite{krengel1977semiamarts,krengel1978semiamarts,samuel1984comparison,hill1981ratio}, matroid, combinatorial, and posted-price views appear in \cite{kleinberg2019matroid,rubinstein2017combinatorial,correa2017posted,correa2021posted,lucier2017economic}; recent advances include \cite{ezra2022prophet,correa2021blind,bubna2023prophet,dutting2020prophet}; see \cite{hill1992survey} for the classical survey. Shape restrictions from reliability theory appear in risk-neutral prophet-type work, but in different roles: reward-tail asymptotics \citep{kennedy1991asymptotic,livanos2025unified}, random-horizon or uncertain-supply models \citep{alijani2020predict,giambartolomei2025iid}, cost-minimization \citep{livanos2024minimization}, and limited-information posted pricing \citep{azar2014prophet}. Here a recentered IFRA condition is used instead to control a lower-tail CVaR coefficient for full-information reward distributions. The underlying zero-origin IFRA definition itself is classical in reliability theory \citep{kochar1987partial,elBarmi2021estimation}, and modern work also uses this cumulative-hazard formulation \citep{bobotas2024preservation}.

\subsection{Organization}
Section~\ref{sec:prelim} records the Conditional Value-at-Risk convention and notation. Section~\ref{sec:pandora} establishes the CVaR Weitzman index reduction for Pandora's box. Section~\ref{sec:prophet} develops the prophet-inequality results: the tight threshold tail-capture guarantee, the two-item impossibility construction, and the recentered IFRA positive result.

\section{CVaR Convention and Notation}
\label{sec:prelim}

Throughout, $\alpha \in (0,1]$ is fixed. For a random reward $Z \in \mathbb{R}$, the \emph{variational form} of CVaR is
\begin{align}
    \cvar_\alpha(Z)
    &=
    \sup_{\lambda \in \R}
    \braces{ \lambda - \tfrac{1}{\alpha} \E{(\lambda - Z)_+} }
    \notag\\
    &=
    \sup_{\lambda \in \R}
    \braces{ \tfrac{1}{\alpha} \E{ \min\braces{Z, \lambda} }
    - \tfrac{1-\alpha}{\alpha} \lambda } ,
    \label{eq:cvar-reward}
\end{align}
where the second equality follows from $(\lambda - Z)_+ = \lambda - \min\{\lambda, Z\}$. The \emph{quantile-integral form} is
\begin{equation}
    \cvar_\alpha(Z) \;=\; \frac{1}{\alpha} \int_0^\alpha Q_Z(u) \, du \ ,
    \label{eq:cvar-quantile}
\end{equation}
where $F_Z$ is the cumulative distribution function (cdf) of $Z$ and $Q_Z(u) = \inf\{z : F_Z(z) \geq u\}$. We write $L_\alpha(Z) \defeq \alpha\, \cvar_\alpha(Z)$ where convenient. At $\alpha = 1$ all forms reduce to $\E{Z}$; as $\alpha \downarrow 0$, $\cvar_\alpha(Z)$ approaches the essential infimum.

%%%%%%%%%%%%%%%%%%%%%%%%%%%%%%%%%%%%%%%%%%%%%%%%%%%%%%%%%%%%%%%%%%%%%%
\section{CVaR Pandora's Box}
\label{sec:pandora}

\noindent\textbf{Setup.}
We are given \(n\) closed boxes. Box \(i\) contains a random prize \(X_i \in \R\) with a known distribution, and may be opened at a known deterministic cost $c_i > 0$; prizes across boxes are independent and integrable. The boxes are inspected sequentially: each round the decision maker may either open any new box (paying its cost and observing its prize) or stop and walk away with the prize from any previously opened box. For any policy, let \(\bbI_i\in\{0,1\}\) indicate whether box \(i\) is opened, and let \(\bbA_i\in\{0,1\}\) indicate whether box \(i\)'s prize is selected. Feasibility requires \(\bbA_i\le \bbI_i\) for all \(i\) and \(\sum_i\bbA_i=1\) almost surely. Thus the selected prize is \(\sum_i\bbA_iX_i\), and the total inspection cost is \(\sum_i\bbI_i c_i\).
% Critically, an opened box may be revisited indefinitely. This freedom makes the commitment relatively soft: information acquired through inspection is not lost when the next box is opened.

\noindent\textbf{Classical objective.}
The classical objective \citep{weitzman1979optimal} is to maximize the expected selected prize minus inspection costs:
\[
    \max_\pi ~
    \E{ \sum_{i=1}^n \bbA_iX_i - \sum_{i=1}^n \bbI_i c_i } ,
\]
where \(\pi\) ranges over all adaptive policies that determine which boxes to open and which opened prize to select. Weitzman's result is that the problem admits an optimal policy of a particularly clean form: assign each box an index $\sigma_i$ that depends only on its own prize distribution and inspection cost, then inspect boxes in decreasing order of $\sigma_i$ and stop as soon as the best prize observed so far exceeds the next index. The index $\sigma_i$ is determined by the indifference condition $c_i = \E{(X_i - \sigma_i)_+}$.

\noindent\textbf{CVaR formulation.}
We now consider the risk-sensitive variant in which the same selected prize is evaluated under CVaR, while the deterministic inspection costs remain in expectation:
\begin{equation}
    \max_\pi ~
    \braces{ \cvar_\alpha\paren{\sum_{i=1}^n \bbA_i X_i}
    - \E{ \sum_{i=1}^n \bbI_i c_i } } \ ,
    \label{eq:cvar-pandora}
\end{equation}
where \(\pi\) ranges over the same feasible adaptive policies.

\begin{remark}[Cost treatment]
We evaluate only the prize under CVaR, not the inspection costs. This reflects the typical applications of Pandora's box: the costs are deterministic or nearly deterministic and modest, while the prizes carry the distributional risk. When $\alpha = 1$, problem \eqref{eq:cvar-pandora} recovers the classical formulation of \cite{weitzman1979optimal}.
\end{remark}

\subsection{Main Result: A Threshold Policy with One-Dimensional Search}

\noindent\textbf{Transformed-prize indices.}
The CVaR variant retains the index structure of the classical problem. The optimal policy is again of Weitzman index form, but with a Gittins-type index \citep{russo2021gittinsucb} computed against a \emph{capped and rescaled} prize distribution. The cap emerges from a single scalar parameter $t$, which is optimized in a separate outer one-dimensional problem.

For each $t \in \R$, define the \emph{$t$-indexed opening threshold} $\sigma_i(t)$ for box $i$ to be the unique solution to
\begin{equation}
    c_i \;=\; \E{ \paren{ \tfrac{1}{\alpha} \min\braces{t, X_i} - \sigma_i(t) }_+ } \ ,
    \label{eq:def-sigma}
\end{equation}
and the \emph{$t$-capped value}
\begin{equation}
    \kappa_i(t) \;\defeq\;
    \min \braces{ \tfrac{1}{\alpha} \min\braces{t, X_i}, ~ \sigma_i(t) } \ .
    \label{eq:def-kappa}
\end{equation}
The existence and uniqueness of $\sigma_i(t)$ follow as follows. For fixed \(t\), define
\(
    h_{i,t}(s)
    \defeq
    \E{\paren{\alpha^{-1}\min\{t,X_i\}-s}_+}.
\)
The function \(h_{i,t}\) is continuous and nonincreasing in \(s\), tends to infinity as \(s\to-\infty\), and is zero for \(s\geq \alpha^{-1}t\). It is strictly decreasing on the region where the expectation is positive, so the assumption \(c_i>0\) gives a unique solution.
% We use the standard finite-instance Weitzman index theorem and value identity for these integrable transformed prizes and deterministic positive inspection costs.

\begin{theorem}[CVaR Weitzman index rule]
\label{thm:pandora-main}
Fix \(\alpha\in(0,1)\), and suppose the one-dimensional objective below attains its maximum. Let
\begin{equation}
    t^\star \;\in\;
    \argmax_{t \in \R} ~ \E{ \max_{i \in [n]} \kappa_i(t) } - \tfrac{1-\alpha}{\alpha}\, t  \ ,
    \label{eq:def-tstar}
\end{equation}
and let $\sigma_i(t^\star)$ be the indices defined by \eqref{eq:def-sigma} with $t = t^\star$. Consider the policy that, at each round, opens the unopened box with the largest index $\sigma_i(t^\star)$ and stops as soon as the largest seen value $x$ satisfies $\tfrac{1}{\alpha} \min\{t^\star, x\}$ exceeds every remaining index, then selects an opened box with largest observed prize. This policy is optimal for \eqref{eq:cvar-pandora}.
\end{theorem}

Theorem \ref{thm:pandora-main} shows that the CVaR objective preserves the architecture of Weitzman's policy: a per-box index, an opening order, and a stopping rule based on observed values. The difference is that the indices are computed on a risk-adjusted reward scale rather than on the raw prizes. Roughly speaking, large realizations are capped before entering the reservation-value calculation, so a box is not ranked highly merely because it has rare extreme upside. Instead, the index measures the box's contribution to the lower-tail value that CVaR emphasizes. The proof below shows how this risk-adjusted scale is obtained from the variational representation of CVaR.

\begin{remark}[Computing \(t^\star\)]
Let \(G(t) \defeq \E{\max_{i \in [n]} \kappa_i(t)} - \tfrac{1-\alpha}{\alpha} t\) denote the outer objective in \eqref{eq:def-tstar}. Under bounded support, \(G\) is eventually decreasing in both tails, so a simple bracketing pass gives a compact search interval. On this interval, the transformed prize \(\alpha^{-1}\min\{t,X_i\}\) is pathwise Lipschitz in \(t\), and the reservation equation implies the same Lipschitz control for \(\sigma_i(t)\) and \(\kappa_i(t)\). Thus \(t^\star\) can be computed by a one-dimensional global search, for example by the Lipschitz-aware method of \cite{malherbe2017global}, once the distributional expectations and indices can be evaluated numerically.
\end{remark}

\subsection{Proof of Theorem~\ref{thm:pandora-main}}

The proof proceeds by a sequence of equivalent reformulations that transform \eqref{eq:cvar-pandora} into an outer maximization over a scalar $t$ of an inner problem that turns out to be a standard Pandora's box instance.

\paragraph{Step 1: Variational form of CVaR.}
Recall from \eqref{eq:cvar-reward} that for any integrable random variable $Y$,
\[
    \cvar_\alpha(Y) \;=\; \sup_{t \in \R} ~
    \braces{ t - \tfrac{1}{\alpha} \E{(t - Y)_+} } \ .
\]
Applying this to $Y = \sum_i \bbA_i X_i$ and substituting into \eqref{eq:cvar-pandora}, we may write
\begin{align}
    \eqref{eq:cvar-pandora}
    &=
    \sup_\pi ~ \sup_{t \in \R}
    \braces{
        t - \tfrac{1}{\alpha}
        \E{ \paren{ t - \sum_{i=1}^n \bbA_i X_i }_+ }
        - \E{\sum_{i=1}^n \bbI_i c_i}
    } .
    \label{eq:pandora-step1}
\end{align}

\paragraph{Step 2: Conversion to a $\min$ form.}
For any reals $t$ and $y$, the algebraic identity $(t - y)_+ = t - \min\{t,y\}$ gives
\[
    t - \tfrac{1}{\alpha} \E{(t - Y)_+}
    \;=\; t - \tfrac{1}{\alpha}\paren{t - \E{\min\{t, Y\}}}
    \;=\; \tfrac{1}{\alpha} \E{\min\{t,Y\}} - \tfrac{1-\alpha}{\alpha} t \ .
\]
Applying this to $Y = \sum_i \bbA_i X_i$ in \eqref{eq:pandora-step1},
\begin{align}
    \eqref{eq:cvar-pandora}
    &=
    \sup_\pi ~ \sup_{t \in \R}
    \braces{
        \tfrac{1}{\alpha}
        \E{ \min\braces{ t, \sum_{i=1}^n \bbA_i X_i } }
        - \tfrac{1-\alpha}{\alpha} t
        - \E{\sum_{i=1}^n \bbI_i c_i}
    } .
    \label{eq:pandora-step2}
\end{align}

\paragraph{Step 3: Exchange of suprema.}
We may exchange the two suprema because both iterated suprema are the supremum of the same objective over all pairs \((\pi,t)\). Thus
\begin{align}
    \eqref{eq:cvar-pandora}
    &=
    \sup_{t \in \R}\sup_\pi
    \braces{
        \tfrac{1}{\alpha}
        \E{ \min\braces{ t, \sum_{i=1}^n \bbA_i X_i } }
        - \tfrac{1-\alpha}{\alpha} t
        - \E{\sum_{i=1}^n \bbI_i c_i}
    } .
    \label{eq:pandora-step3}
\end{align}

\paragraph{Step 4: One-hot reduction.}
The selection vector $\bbA = (\bbA_1, \dots, \bbA_n)$ is one-hot: $\bbA_i \in \{0,1\}$ and $\sum_i \bbA_i = 1$. Hence on every sample path there is a unique index $j$ with $\bbA_j = 1$, and $\sum_i \bbA_i X_i = X_j$. For any function $f: \R \to \R$,
\begin{equation}
    f\paren{\sum_{i=1}^n \bbA_i X_i}
    \;=\; f(X_j)
    \;=\; \sum_{i=1}^n \bbA_i\, f(X_i) \ ,
    \label{eq:one-hot}
\end{equation}
the last equality because exactly one $\bbA_i$ equals $1$. Taking $f(y) = \min\{t, y\}$, we may pass the minimum inside the linear combination:
\[
    \min\braces{t, \sum_{i=1}^n \bbA_i X_i}
    \;=\; \sum_{i=1}^n \bbA_i \min\{t, X_i\} \ .
\]
Substituting this identity into the inner problem in \eqref{eq:pandora-step3},
\begin{align}
    \eqref{eq:cvar-pandora}
    &=
    \sup_{t \in \R}
    \braces{
        \sup_\pi
        \E{ \sum_{i=1}^n \bbA_i \cdot \tfrac{1}{\alpha} \min\{t, X_i\}
        - \sum_{i=1}^n \bbI_i c_i }
        - \tfrac{1-\alpha}{\alpha} t
    } .
    \label{eq:pandora-step4}
\end{align}

\paragraph{Step 5: The inner problem is a classical Pandora instance.}
We now fix $t \in \R$ and consider the inner supremum in \eqref{eq:pandora-step4}. This is precisely the value of the \emph{classical} (risk-neutral) Pandora's box problem with modified prizes $\tilde X_i \defeq \tfrac{1}{\alpha} \min\{t, X_i\}$ and unchanged costs $c_i$. Indeed, the policy chooses some set of boxes to inspect and selects one opened box to claim, and seeks to maximize the expected net reward $\E{\tilde X_{j^\star} - \sum_{i \in I} c_i}$.

By Weitzman's value identity, this inner problem admits an optimal policy of Weitzman index form, where the index $\sigma_i(t)$ for box $i$ is the unique solution to
\[
    c_i \;=\; \E{ (\tilde X_i - \sigma_i(t))_+ }
    \;=\; \E{ \paren{ \tfrac{1}{\alpha} \min\{t, X_i\} - \sigma_i(t) }_+ } \ ,
\]
recovering \eqref{eq:def-sigma}. Moreover, the value of the inner problem is characterized by Weitzman's identity in terms of the \emph{capped} indices: defining $\kappa_i(t) = \min\{\tilde X_i, \sigma_i(t)\}$ as in \eqref{eq:def-kappa},
\begin{equation}
    \sup_\pi
    \E{ \sum_{i=1}^n \bbA_i \cdot \tfrac{1}{\alpha} \min\{t, X_i\}
    - \sum_{i=1}^n \bbI_i c_i }
    =
    \E{ \max_{i \in [n]} \kappa_i(t) } \ .
    \label{eq:weitzman-identity}
\end{equation}

\paragraph{Step 6: Conclusion.}
Combining \eqref{eq:weitzman-identity} with \eqref{eq:pandora-step4},
\begin{equation}
    \eqref{eq:cvar-pandora}
    \;=\;
    \sup_{t \in \R} ~
    \braces{ \E{ \max_{i \in [n]} \kappa_i(t) } - \tfrac{1-\alpha}{\alpha} t } \ .
    \label{eq:pandora-final}
\end{equation}
By the theorem's assumption that the outer objective in \eqref{eq:def-tstar} attains its maximum, choose \(t^\star\) in its argmax. Let \(\pi^\star\) be the Weitzman policy for the transformed instance at \(t^\star\). For this policy, the variational representation used in Steps 1 and 2 says that the original CVaR objective is the supremum over the variational scalar, so it is at least its value at \(t^\star\). Because \(\pi^\star\) attains the inner supremum in \eqref{eq:pandora-step4} at \(t^\star\), this value equals the objective value in \eqref{eq:pandora-final} at \(t^\star\). Since \(t^\star\) attains the supremum in \eqref{eq:pandora-final}, \(\pi^\star\) is optimal. This is the policy stated in the theorem.

%%%%%%%%%%%%%%%%%%%%%%%%%%%%%%%%%%%%%%%%%%%%%%%%%%%%%%%%%%%%%%%%%%%%%%
\section{CVaR Prophet Inequality}
\label{sec:prophet}

\noindent\textbf{Setup and results.}
In the classical prophet inequality, a decision maker observes independent nonnegative rewards \(X_1,\ldots,X_n\) sequentially and must stop irrevocably at one observation, receiving payoff \(S\) (or zero if she never stops). 
The distribution of each \(X_i\) is known, but the arrival order is not. 
With \(M\defeq\max_i X_i\), the goal is to design an online stopping rule whose expected payoff is a constant fraction of the prophet's benchmark \(\E{M}\), uniformly over all independent reward distributions. 
The sharp constant is \(\tfrac{1}{2}\) \citep{krengel1977semiamarts,krengel1978semiamarts}, and it can be achieved by, e.g., a threshold policy at a median of \(M\) \citep{samuel1984comparison}. 

In the CVaR setting, an analogous and natural question is whether an online policy can guarantee a constant fraction of the prophet's CVaR benchmark: for a fixed \(\alpha\in(0,1)\), can \(\cvar_\alpha(S)\) be bounded below by a positive constant, depending only on \(\alpha\), times \(\cvar_\alpha(M)\)? 
Namely, it is the classical distribution-free prophet guarantee with expectation replaced by a lower-tail risk functional.
In this section, we will show in Theorem~\ref{thm:prophet-main} that the answer is negative at this level of generality: on the full class of independent nonnegative reward distributions, no such constant exists.
Instead, the sharp guarantee is an \textit{instance-dependent} coefficient \(B_\alpha(M)\), attained by some threshold policy. 
Section~\ref{sec:draft-ifra-reorganization} shows that, under continuous recentered increasing-failure-rate-average (IFRA) distributions, threshold policies recover an explicit multiplicative \(\alpha\)-only constant approximation guarantee.

\noindent\textbf{Threshold tie-breaking convention.}
A threshold policy fixes a level \(\tau\) before the observations arrive and stops at the first observation strictly above \(\tau\); if no observation is accepted, the payoff is zero. Its no-trigger probability is the probability of reaching the end without stopping. At atoms, we allow independent tie-breaking when an observation equals \(\tau\); by varying these tie-breaking probabilities, the no-trigger probability can be chosen anywhere in \(\brackets{\pr{M<\tau},\pr{M\leq\tau}}\), which simply reduces to \(\pr{M<\tau}\) when the distributions free of atoms.

\noindent\textbf{Tail-capture coefficient.}
Having threshold policies in mind, we now proceed to introduce the tight constant coefficient \(B_\alpha(M)\).
% The coefficient below packages the proof template before the formal argument appears. 
% The numerator has three pieces: the threshold payoff \(\tau(1-r)\), the capped excess payoff \(r\,\E{(M\wedge t-\tau)_+}\), and the CVaR variational penalty \((1-\alpha)t\). 
We define the \emph{threshold tail-capture coefficient} to be 
\[
    B_\alpha(M)
    \;\defeq\;
    \sup_{\tau\geq0} ~
    \sup_{t\geq\tau} ~
    \sup_{\pr{M<\tau}\leq r\leq \pr{M\leq\tau}} ~
    \frac{
        \paren{1-r} \tau
        +r\,\E{(M\wedge t-\tau)_+}
        -(1-\alpha)t
    }{L_\alpha(M)} .
\]
It optimizes over a threshold \(\tau\), a cap \(t\ge\tau\), and a no-trigger probability \(r\) obtained by tie-breaking at the threshold. 
Thus, whenever the supremum is attained, the threshold policy realizing the coefficient is read directly from an optimizer \((\tau^\star,t^\star,r^\star)\): use \(\tau^\star\) as the stopping threshold and choose equality tie-breaking so that the no-trigger probability is \(r^\star\); the cap \(t^\star\) certifies the CVaR lower bound but is not part of the stopping rule.
Intuitively, this coefficient asks whether some threshold level and cap together capture a nontrivial portion of the prophet's lower $\alpha$-tail value $L_\alpha(M)$.
% Theorem~\ref{thm:prophet-main} says that a threshold policy achieves this coefficient, up to an arbitrarily small loss, and this leading constant is worst-case optimal against all online policies.
% The no-constant impossibility for fixed $\alpha<1$ follows from the same hard instance used for tightness.

\subsection{A Tight Instance-Dependent Multiplicative Guarantee}

\begin{theorem}[Tight threshold tail-capture guarantee]
\label{thm:prophet-main}
\label{thm:prophet-tail-capture}
Assume that $0<L_\alpha(M)<\infty$.
Fix $\alpha\in(0,1)$. For every prophet inequality instance with independent nonnegative reward distributions and every  arrival order fixed in advance, for every $\eta>0$ there exists a threshold policy $S_\tau$ such that
\[
    \frac{\cvar_\alpha(S_\tau)}{\cvar_\alpha(M)}
    \;\geq\; B_\alpha(M)-\eta .
\]
Moreover, the leading constant is worst-case tight: for every $c>1$, there exists a two-item instance such that every online stopping rule $S$ satisfies
\[
    \frac{\cvar_\alpha(S)}{\cvar_\alpha(M)}
    \;\leq\; c\,B_\alpha(M) .
\]
\end{theorem}

\begin{remark}[Risk-neutral as a special case]
Under the lower-tail convention used here, $\alpha=1$ gives $\cvar_1(Z)=\E{Z}$, so the objective becomes the classical risk-neutral prophet objective. Although Theorem~\ref{thm:prophet-main} is stated for $\alpha<1$, its positive part and the definition of $B_\alpha(M)$ extend to $\alpha=1$ provided that \(0<\E{M}<\infty\). We can recover the Samuel--Cahn median-threshold guarantee. To see this, we take a median threshold \(\tau\) for \(M\), so \(\pr{M<\tau}\leq\tfrac12\leq\pr{M\leq\tau}\), and choose the tie-breaking probabilities so that the no-trigger probability is exactly \(\tfrac12\). Letting \(t\to\infty\) and using monotone convergence then gives
\[
    B_1(M)
    \;\geq\;
    \frac{\tfrac{1}{2}\tau
    + \tfrac{1}{2}\E{(M-\tau)_+}}{\E{M}}
    \;\geq\; \frac{1}{2},
\]
where the second inequality follows because \(M\leq \tau+(M-\tau)_+\).
% Thus the same coefficient that governs the lower-tail CVaR problem also contains the classical \(1/2\) endpoint.
\end{remark}

\begin{myproof}
\emph{Step 1: Positive guarantee.}
First note that \(0\leq B_\alpha(M) \). This can be seen by taking \(\tau=0\) and \(t=0\) in the definition. 
If $B_\alpha(M)=0$, the claim is immediate because CVaR ratios are nonnegative. 
Otherwise, we choose $\tau\geq0$, $t\geq\tau$, and $r\in\brackets{\pr{M<\tau},\pr{M\leq\tau}}$ such that
\begin{align}
        \tau\paren{1-r}
        +r\,\E{(M\wedge t-\tau)_+}
        -(1-\alpha)t
    \geq
    \paren{B_\alpha(M)-\eta}L_\alpha(M).
    \label{eq:tail-capture-near-opt}
\end{align}
Consider the threshold policy with tie-breaking chosen so that the no-trigger probability is $r$.
Let \(X_i\) denote the \(i\)th arriving observation in the fixed arrival order.
Let \(A_i\) be the event that the policy accepts at round \(i\) for $i=1,\ldots,n$, and let \(A_0\) be the event that the policy never accepts.  
For each \(i\), let \(B_i\) be the event 
\[
    B_i
    =
    \{X_i>\tau\}
    \cup
    \bigl(\{X_i=\tau\}\cap\{\text{the tie at }i\text{ is accepted}\}\bigr).
\]
Let \(r_i\defeq\pr{B_i^c}\). Then the chosen tie-breaking probabilities satisfy \(\prod_{i=1}^{n} r_i=r\).
Finally, let \(C_i\defeq\bigcap_{j<i}B_j^c\) be the event that no earlier observation has been accepted before observation \(i\). Then \(A_i=B_i\cap C_i\), where \(C_i\) is independent of \(X_i\) and the tie-breaking at observation \(i\). Therefore,
\[
    \pr{A_i}
    =
    \paren{\prod_{j<i} r_j}\paren{1-r_i},
    \qquad
    \pr{A_0}=r.
\]
For any real value \(a\), \(a=\min\{a,\tau\}+(a-\tau)_+\). With \(a=X_i\wedge t\) and \(t\geq\tau\), this gives \(X_i\wedge t=\tau+(X_i\wedge t-\tau)_+\) on \(\{X_i\geq\tau\}\). Since \(A_i\subseteq\{X_i\geq\tau\}\), we have \((X_i\wedge t)\one{A_i}=\tau\one{A_i}+(X_i\wedge t-\tau)_+\one{A_i}\).
Given that $\sum_{i=0}^n \one{A_i}=1$, we have for $t \geq \tau$, 
\begin{align}
    \E{S_\tau\wedge t}
    &= \sum_{i=1}^{n} \brackets{ \tau \pr{A_i} + \E{ (X_i\wedge t-\tau)_+\one{A_i} } } \notag \\
    &= 
    \tau\sum_{i=1}^n \pr{A_i}
    + \sum_{i=1}^n \E{(X_i\wedge t-\tau)_+\one{B_i}\one{C_i}} \notag\\
    &=
    \tau(1-r)
    + \sum_{i=1}^n \E{(X_i\wedge t-\tau)_+\one{B_i}}
        \prod_{j<i}r_j \label{eq:uses_C_ind_B}\\
    &=
    \tau(1-r)
    + \sum_{i=1}^n \E{(X_i\wedge t-\tau)_+}
        \prod_{j<i}r_j
    \label{eq:threshold-cap-drop-trigger}\\
    &\geq
    \tau(1-r)
    + r\sum_{i=1}^n \E{(X_i\wedge t-\tau)_+}
    \label{eq:threshold-cap-reach}\\
    &\geq
    \tau\paren{1-r}
    + r\,\E{(M\wedge t-\tau)_+}.
\label{eq:threshold-cap-bound}
\end{align}
Equation~\eqref{eq:uses_C_ind_B} uses the independence of \(C_i\) from \(X_i\) and the tie-breaking at observation. 
Equation~\eqref{eq:threshold-cap-drop-trigger} is true due to that on \(B_i^c\), \(X_i\leq\tau\), so \((X_i\wedge t-\tau)_+=0\).
The inequality in \eqref{eq:threshold-cap-reach} follows from
\(
    \prod_{j<i}r_j
    \;\geq\; \prod_{j=1}^n r_j
    \;=\; r
\)
and the inequality in \eqref{eq:threshold-cap-bound} follows from the pointwise bound
\[
    \sum_{i=1}^n (X_i\wedge t-\tau)_+
    \;
    \geq \max_i  ~ (X_i\wedge t-\tau)_+
    =\; \paren{\max_i (X_i\wedge t)-\tau}_+
    \;=\; (M\wedge t-\tau)_+ .
\]
Recall the capped variational form for nonnegative \(Y\),
\[
    L_\alpha(Y)
    =
    \sup_{u\geq0}\braces{\E{Y\wedge u}-(1-\alpha)u}
    \qquad\text{(cf. \eqref{eq:cvar-reward}).}
\]
Applying this identity to \(Y=S_\tau\) and choosing the feasible cap \(u=t\), then combining with \eqref{eq:threshold-cap-bound}, gives
\begin{align}
    L_\alpha(S_\tau)
    &\geq \E{S_\tau\wedge t}-(1-\alpha)t \\
    &\geq
    \tau\paren{1-r}
    + r\,\E{(M\wedge t-\tau)_+}
    -(1-\alpha)t . \label{eq:L_alpha_lower_bound}
\end{align}
We note in passing that this calculation shows that the numerator in the definition of \(B_\alpha(M)\) is a lower bound on \(L_\alpha(S_\tau)\) for any feasible \((\tau,t,r)\), and therefore, $B_\alpha(M) \leq 1$. 
% The lower bound follows by taking \(\tau=t=0\) and \(r=0\). For the upper bound, any feasible \((\tau,t,r)\) can be realized by the threshold tie-breaking convention; the calculation below gives its numerator as a lower bound on \(L_\alpha(S_\tau)\). Since \(S_\tau\leq M\) almost surely and \(L_\alpha\) is monotone, this numerator is at most \(L_\alpha(M)\). 

Dividing \eqref{eq:L_alpha_lower_bound} by \(L_\alpha(M)\) and then applying \eqref{eq:tail-capture-near-opt} gives
\[
    \frac{L_\alpha(S_\tau)}{L_\alpha(M)}
    \geq B_\alpha(M)-\eta .
\]
Since $L_\alpha(Y)=\alpha\,\cvar_\alpha(Y)$, this proves the positive guarantee.

\emph{Step 2: Tightness.}
We construct an instance with two items, where the items arrive in the fixed order $X_1$ then $X_2$. 
Fix $\rho\in(0,1)$ and an integer $K\geq1$.
The exact values of $\rho$ and $K$ depends on $c$ and will be specified later.
Define
\begin{equation}
    p_j \;\defeq\; \frac{(1-\rho)\rho^{j-1}}{1-\rho^K},
    \qquad
    v_j \;\defeq\; \frac{1}{p_j},
    \qquad j=1,\ldots,K .
    \label{eq:hard-pj}
\end{equation}
Since \(\rho\in(0,1)\), the probabilities \(p_j\) decrease with \(j\), and therefore the values \(v_j=1/p_j\) increase with \(j\). 
Let $X_1$ take value $v_j$ with probability $p_j$. Let $X_2$ be independent of $X_1$, with $\pr{X_2=0}=\alpha$ and $\pr{X_2=H}=1-\alpha$ for some $H>v_K$. 
Since $H>v_K$,
\[
    M =
    \begin{cases}
        H, & X_2=H,\\
        X_1, & X_2=0.
    \end{cases}
\]
The lower $\alpha$-tail of $M$ is exactly the event $\{X_2=0\}$, so regardless of $\rho$,
\begin{align}
    \cvar_\alpha(M)
    &= \E{ M \mid X_2=0}
    = \E{X_1}
    =
    \sum_{j=1}^K p_jv_j
    =
    K,
    \notag\\
    L_\alpha(M)
    &=
    \alpha K .
    \label{eq:tail-capture-prophet-value}
\end{align}

The choice $\tau=t=v_1$ and $r=0$ is feasible in the definition of $B_\alpha(M)$, since $\pr{M<v_1}=0$. Substituting this feasible triple into the definition of \(B_\alpha(M)\) gives
\[
    B_\alpha(M)
    \geq
    \frac{
        v_1(1-0)
        +0\cdot\E{(M\wedge v_1-v_1)_+}
        -(1-\alpha)v_1
    }{L_\alpha(M)}
    =
    \frac{\alpha v_1}{L_\alpha(M)} .
\]
Combining this with \eqref{eq:tail-capture-prophet-value} gives
\begin{equation}
    B_\alpha(M)
    \;\geq\; \frac{v_1}{K}
    \;=\; \frac{1-\rho^K}{(1-\rho)K}.
    \label{eq:tail-capture-B-hard}
\end{equation}
We will use this lower bound below. 

Now consider an arbitrary online policy. 
Without loss of generality, we consider policies that will accept the second item, if it passes the first item. 
Let $s_j\in[0,1]$ be the conditional probability that the policy accepts the first item after observing the event $X_1=v_j$, and let $S$ denote the final payoff of this policy. 
Then
\[
    \pr{S=0}=\alpha \cdot \paren{1-\sum_{j=1}^K p_js_j},\qquad
    \pr{S=v_j}=p_js_j, ~ j=1,\ldots,K, \qquad
    \pr{S=H}=(1-\alpha)\paren{1-\sum_{j=1}^K p_js_j}.
 \]
The lower $\alpha$-tail of $S$ never reaches $H$, since the mass at or below $v_K$ is \(\alpha\paren{1-\sum_{j=1}^K p_js_j}+\sum_{j=1}^K p_js_j\geq\alpha\).

If \(\sum_{j=1}^K p_js_j=0\), then $\cvar_\alpha(S)=0$, and there is nothing to prove. 
Otherwise, we proceed to calculate $\cvar_\alpha(S)$.  To this end, let $m$ be the smallest index such that \(\sum_{j=1}^m p_js_j\geq\alpha\sum_{j=1}^K p_js_j\).  
The lower $\alpha$-tail of $S$ first includes the zero mass \(\alpha\paren{1-\sum_{j=1}^K p_js_j}\), which contributes nothing to \(\alpha\cvar_\alpha(S)\). The remaining tail mass is \(\alpha\sum_{j=1}^K p_js_j\). By minimality of \(m\), \(\sum_{j=1}^{m-1}p_js_j<\alpha\sum_{j=1}^K p_js_j\leq\sum_{j=1}^{m}p_js_j\), so this remaining mass contains all levels $1,\ldots,m-1$ and exactly \(\alpha\sum_{j=1}^K p_js_j-\sum_{j=1}^{m-1}p_js_j\) units of mass at \(v_m\). Hence
\begin{equation}
    \cvar_\alpha(S)
    = \frac{1}{\alpha} \cdot \paren{ 
    \sum_{j=1}^{m-1}p_js_jv_j
    +
    \paren{\alpha\sum_{j=1}^K p_js_j-\sum_{j=1}^{m-1}p_js_j}v_m } .
    \label{eq:tail-decomp}
\end{equation}
We proceed to upper bound $\cvar_\alpha(S)$ in terms of a function of $\rho$ and $\alpha$. 
If $m=1$, the first term is zero. If $m\geq2$, then $v_j\leq v_{m-1}=1/p_{m-1}$ for $j\leq m-1$, so
\[
    \sum_{j=1}^{m-1}p_js_jv_j
    \leq \frac{1}{p_{m-1}}\sum_{j=1}^{m-1}p_js_j .
\]
For \(m\geq2\), minimality of $m$ implies that 
\[
    \sum_{j=1}^{m-1}p_js_j
    < \alpha\sum_{j=1}^K p_js_j
    =
    \alpha\sum_{j=1}^{m-1}p_js_j
    + \alpha\sum_{j=m}^K p_js_j,
\]
and rearranging yields
\[
    (1-\alpha)\sum_{j=1}^{m-1}p_js_j
    < \alpha\sum_{j=m}^K p_js_j
    \leq \alpha\sum_{j=m}^K p_j
    \leq \frac{\alpha\rho p_{m-1}}{1-\rho},
\]
where the last inequality uses the geometric tail: for \(j\geq m\), \(p_j=p_{m-1}\rho^{j-m+1}\), so
$
\sum_{j=m}^K p_j
= p_{m-1} \sum_{\ell=1}^{ K-m+1} \rho^\ell
\leq p_{m-1} \sum_{\ell=1}^{\infty} \rho^\ell
= \frac{\rho p_{m-1}}{1-\rho}.
$

Together with the trivial \(m=1\) case, this implies that, for every $m$,
\begin{equation}
    \sum_{j=1}^{m-1}p_js_jv_j
    \leq \frac{\alpha\rho}{(1-\alpha)(1-\rho)} .
    \label{eq:full-level-final}
\end{equation}
Similarly,
\[
    \alpha\sum_{j=1}^K p_js_j-\sum_{j=1}^{m-1}p_js_j
    \leq \alpha\sum_{j=m}^K p_js_j
    \leq \frac{\alpha p_m}{1-\rho},
\]
and since $v_m=1/p_m$,
\begin{equation}
    \paren{\alpha\sum_{j=1}^K p_js_j-\sum_{j=1}^{m-1}p_js_j}v_m
    \leq \frac{\alpha}{1-\rho}.
    \label{eq:boundary-final}
\end{equation}
Finally, substituting \eqref{eq:full-level-final} and \eqref{eq:boundary-final} into \eqref{eq:tail-decomp} gives an upper bound 
\begin{equation}
    \alpha\cvar_\alpha(S)
    \leq
    \frac{\alpha}{1-\rho}
    \paren{1+\frac{\rho}{1-\alpha}} .
    \label{eq:tail-capture-policy-upper}
\end{equation}
Dividing \eqref{eq:tail-capture-policy-upper} by \eqref{eq:tail-capture-prophet-value} gives \eqref{eq:tail-capture-ratio-upper}. Equation~\eqref{eq:tail-capture-factorization} only multiplies and divides by \(1-\rho^K\), and \eqref{eq:tail-capture-tight-factor} uses \eqref{eq:tail-capture-B-hard}:
\begin{align}
    \frac{\cvar_\alpha(S)}{\cvar_\alpha(M)}
    &\leq
    \frac{1+\rho/(1-\alpha)}{(1-\rho)K}
    \label{eq:tail-capture-ratio-upper}\\
    &=
    \frac{1+\rho/(1-\alpha)}{1-\rho^K}
    \cdot
    \frac{1-\rho^K}{(1-\rho)K}
    \label{eq:tail-capture-factorization}\\
    &\leq
    \frac{1+\rho/(1-\alpha)}{1-\rho^K}\,B_\alpha(M).
    \label{eq:tail-capture-tight-factor}
\end{align}
Given any \(c>1\), we shall choose \(\rho \in (0, 1)\) small enough so that
\(
    1+\rho/(1-\alpha) < c .
\)
Moreover, we shall choose sufficiently large \(K\), such that
\[
    \frac{1+\rho/(1-\alpha)}{1-\rho^K} < c .
\]
Namely, the multiplicative factor before $B_\alpha(M)$ in \eqref{eq:tail-capture-tight-factor} can therefore be made smaller than \(c\). This proves tightness.
\end{myproof}

\begin{corollary}[No uniform constant approximation guarantee without distributional assumptions]
\label{thm:prophet-impossibility}
Fix any $\alpha \in (0,1)$ and $\varepsilon > 0$. There exists a two-observation instance with independent nonnegative reward distributions and an arrival order fixed in advance such that every possibly randomized stopping rule $S$ satisfies
\[
    \frac{\cvar_\alpha(S)}{\cvar_\alpha(M)} \;\leq\; \varepsilon .
\]
% Consequently,
% \[
%     \inf_{\substack{\text{independent nontrivial instance}\\
%     \text{with fixed ex-ante order}}} ~
%     \sup_{\text{possibly randomized }S}
%     \frac{\cvar_\alpha(S)}{\cvar_\alpha(M)} \;=\; 0 .
% \]
\end{corollary}

\begin{myproof}
Use the construction in Step 2 of Theorem~\ref{thm:prophet-main} with any fixed $\rho\in(0,1)$. The direct ratio bound \eqref{eq:tail-capture-ratio-upper} gives, for every online policy,
\(
    \frac{\cvar_\alpha(S)}{\cvar_\alpha(M)}
    \leq
    \frac{1+\rho/(1-\alpha)}{(1-\rho)K}.
\)
Taking $K$ large enough makes the right-hand side at most $\varepsilon$.
% The factor \(1-\rho^K\) appears only in the tightness proof after rewriting this ratio relative to \(B_\alpha(M)\).
\end{myproof}

\begin{remark}[Source of impossibility]
\label{rem:impossibility-comment}
The two-item construction makes the obstruction transparent. The second item is either zero, with probability $\alpha$, or very large, with probability $1-\alpha$. Hence the prophet's lower $\alpha$-tail is exactly the branch in which the second item is zero, and on that branch the prophet receives $X_1$. An online policy, however, must decide whether to accept the realized value of $X_1$ before learning which branch of $X_2$ will occur. Rejecting $X_1$ preserves the chance of receiving the large value $H$, which is useful \textit{in expectation}, but that upside lies \textit{outside} the lower $\alpha$-tail and therefore does not improve CVaR. By spreading $X_1$ over many increasingly rare and increasingly large values, the construction makes the prophet's CVaR grow in an unbounded manner while every online policy's CVaR remains bounded.
\end{remark}

%%%%%%%%%%%%%%%%%%%%%%%%%%%%%%%%%%%%%%%%%%%%%%%%%%%%%%%%%%%%%%%%%%%%%%
\subsection{A Constant Prophet Guarantee under Recentered IFRA}
\label{sec:draft-ifra-reorganization}

In this subsection, we show that a constant-factor prophet guarantee, depending only on \(\alpha\), can be recovered once the reward distributions satisfy a recentered IFRA condition. The two-item construction in Remark~\ref{rem:impossibility-comment} works by separating the payoff that makes the prophet large from the lower tail that CVaR evaluates. Recentered IFRA prevents this separation from being arbitrary: large rewards cannot be detached from the lower quantiles of the maximum \(M\).
This lets threshold policies be certified in two complementary ways: either the policy stops often enough at a sufficiently high threshold, or the values above the threshold are still tied closely enough to the prophet's lower tail that the same policy recovers additional value from those accepted rewards.

\subsubsection{Recentered IFRA Distributions}

We use the following continuous, left-endpoint recentered version of the classical increasing-failure-rate-average (IFRA) condition in reliability theory.

\begin{definition}[Continuous recentered IFRA distributions]
\label{def:draft-ifra}
Let \(X\) be a nonnegative random variable with cumulative distribution function \(F\). We say that \(X\) has a \emph{continuous recentered increasing-failure-rate-average distribution}, if there is an interval \([s,u)\), with \(0\le s<u\le\infty\), such that:
\begin{enumerate}
    \item[(i)] \(X\) is atomless and absolutely continuous, with positive continuous density on \((s,u)\);
    \item[(ii)] \(F(x)=0\) for \(x\le s\), and, when \(u<\infty\), \(F(x)=1\) for \(x\ge u\);
    \item[(iii)] the cumulative hazard \(H(x)\defeq-\log(1-F(x))\) satisfies that \(x\mapsto H(x)/(x-s)\) is nondecreasing on \((s,u)\).
\end{enumerate}
\end{definition}

Reliability theory usually defines the increasing-failure-rate-average condition for a nonnegative lifetime \(Y\) whose support starts at zero: with cumulative hazard \(H_Y\), the average cumulative hazard \(y\mapsto H_Y(y)/y\) is nondecreasing \citep{kochar1987partial,elBarmi2021estimation}. This is the standard cumulative-hazard formulation; see also \citet{bobotas2024preservation}. Definition~\ref{def:draft-ifra} applies the same condition after translating the left endpoint of the support to zero. Indeed, if \(Y=X-s\), then \(H_Y(y)=H(s+y)\), so condition (iii) is exactly the usual IFRA condition for the recentered variable \(Y\).
%  When \(s=0\), this reduces to the usual IFRA condition for \(X\); when \(s>0\), it differs from requiring \(x\mapsto H(x)/x\) to be nondecreasing on the original scale.

\begin{remark}[Examples and scope of recentered IFRA]
The class includes many standard distributions. Let \(h(x)=H'(x)\) denote the hazard rate on \((s,u)\). If \(h\) is nondecreasing, then \(H(x)=\int_s^x h(z)\,dz\) is convex and satisfies \(H(s)=0\). Hence, for any \(s<x_1<x_2<u\), convexity gives
\[
    \frac{H(x_1)-H(s)}{x_1-s}
    \le
    \frac{H(x_2)-H(s)}{x_2-s}.
\]
Since \(H(s)=0\), this is exactly \(H(x_1)/(x_1-s)\le H(x_2)/(x_2-s)\). Thus every distribution satisfying the support and regularity conditions in Definition~\ref{def:draft-ifra} and having nondecreasing hazard rate on \((s,u)\) is recentered IFRA; see \citet{barlow1975statistical,barlow1996mathematical}. A convenient sufficient condition for this monotone-hazard-rate property is log-concavity of the density \citep{bagnoli2005log}. Consequently, after shifting the left support endpoint to \(s\), Definition~\ref{def:draft-ifra} includes the standard log-concave families satisfying its support and regularity conditions: exponential distributions, uniform distributions on a finite interval, Weibull and Gamma distributions with shape parameter at least one, Erlang distributions as the integer-shape Gamma subfamily, chi-square distributions with at least two degrees of freedom, and normal distributions truncated to an interval \([s,u]\) with \(0\le s<u\le\infty\) \citep{bagnoli2005log}.
\end{remark}

\subsubsection{A Constant Multiplicative Prophet Guarantee}

Fix \(\alpha\in(0,1]\), set \(M\defeq\max_i X_i\), and let \(F_M\) and \(Q\) denote the cumulative distribution function and quantile function of \(M\). Write
\[
    \ell(u)\defeq-\log(1-u),
    \qquad
    \Psi(a)\defeq\int_0^a\ell(u)\,du,
    \qquad
    \Psi(1)=1,
\]
where \(0\le u,a<1\) before the continuous extension.
The constant in Theorem~\ref{thm:draft-prophet-ifra} is the larger of two lower bounds, each coming from a different way of evaluating the same class of threshold policies. The \emph{threshold-floor bound} uses only the payoff guaranteed at the threshold itself. Let \(q_\alpha\in(0,\alpha)\) be the unique solution of \(\ell(q_\alpha)=(\alpha-q_\alpha)/(1-q_\alpha)\), and define
\[
    \rho_{\mathrm{floor}}(\alpha)
    \defeq
    \left[
        \log\frac{\alpha}{\alpha-q_\alpha}
        +
        \frac{\Psi(\alpha)-\Psi(q_\alpha)}
        {(\alpha-q_\alpha)\ell(q_\alpha)}
    \right]^{-1}.
\]
The \emph{tail-recovery bound} uses the same threshold policy but evaluates it at a cap above the threshold, thereby recovering additional lower-tail value from the prophet's residual tail. Define the admissible quantile levels by
\[
    \mathcal D_\alpha
    \defeq
    \{q\in(0,\alpha)\colon q\le 1/2,\; q(1-q)\ge1-\alpha\}.
\]
For \(q\in\mathcal D_\alpha\), let
\[
    G_{\mathrm{tail}}(\alpha,q)
    \defeq
    (1-2q)\frac{\ell(q)}{\Psi(\alpha)+\ell(q)-q}
    +q\,\frac{\Psi(1-(1-\alpha)/q)}{\Psi(\alpha)}.
\]
Finally,
\[
    \rho_{\mathrm{tail}}(\alpha)
    \defeq
    \begin{cases}
    \sup_{q\in\mathcal D_\alpha}G_{\mathrm{tail}}(\alpha,q),
    &\mathcal D_\alpha\ne\emptyset,\\
    0,&\mathcal D_\alpha=\emptyset,
    \end{cases}
    \qquad
    \rho(\alpha)
    \defeq
    \max\{\rho_{\mathrm{floor}}(\alpha),\rho_{\mathrm{tail}}(\alpha)\}.
\]

\begin{assumption}[Continuous recentered IFRA prophet instance]
\label{assump:draft-ifra-prophet}
The rewards \(X_1,\ldots,X_n\) are independent and nonnegative, each \(X_i\) satisfies Definition~\ref{def:draft-ifra}, the arrival order is fixed in advance and value-oblivious, and \(0<L_\alpha(M)<\infty\).
\end{assumption}

Under Assumption~\ref{assump:draft-ifra-prophet}, if \(F_i\) denotes the cumulative distribution function of \(X_i\), then \(F_M(x)=\prod_i F_i(x)\). In particular, \(M\) is atomless, and \(F_M(Q(q))=q\) for every \(q\in(0,1)\).

\begin{theorem}[Threshold-policy bound under continuous recentered IFRA]
\label{thm:draft-prophet-ifra}
Under Assumption~\ref{assump:draft-ifra-prophet}, for every \(\varepsilon>0\), there exists a threshold policy \(S_\tau\) such that
\[
    \frac{\cvar_\alpha(S_\tau)}{\cvar_\alpha(M)}
    =
    \frac{L_\alpha(S_\tau)}{L_\alpha(M)}
    \ge
    \rho(\alpha)-\varepsilon .
\]
\end{theorem}

\begin{figure}[t]
    \centering
    \includegraphics[width=0.78\textwidth]{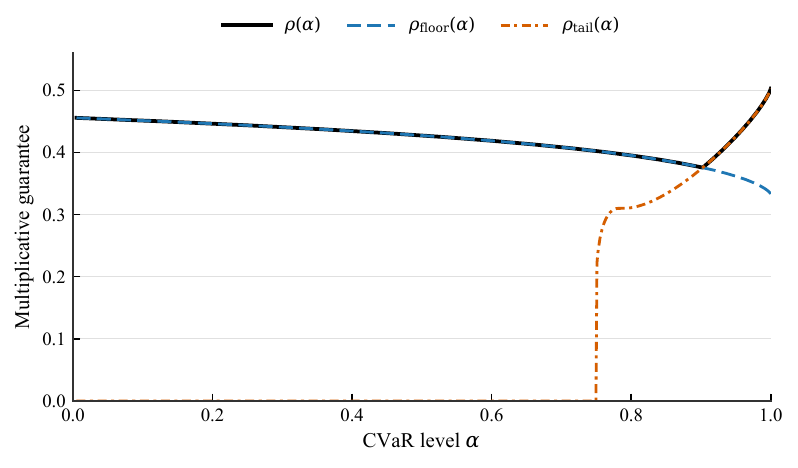}
    \caption{The IFRA bound \(\rho(\alpha)=\max\{\rho_{\mathrm{floor}}(\alpha), \rho_{\mathrm{tail}}(\alpha)\}\) for \(0<\alpha\le1\). The solid curve is \(\rho\), the dashed curve is the threshold-floor bound \(\rho_{\mathrm{floor}}\), and the dash-dotted curve is the tail-recovery bound \(\rho_{\mathrm{tail}}\), computed by numerically optimizing \(G_{\mathrm{tail}}(\alpha,q)\) over \(q\in\mathcal D_\alpha\). The tail-recovery bound is active near the risk-neutral endpoint; at \(\alpha=1\), the feasible choice \(q=1/2\) gives \(G_{\mathrm{tail}}(1,1/2)=1/2\), while the optimized endpoint value is \(\rho_{\mathrm{tail}}(1)\approx0.503\).}
    \label{fig:draft-ifra-bound}
\end{figure}

Figure~\ref{fig:draft-ifra-bound} plots the threshold-floor bound, the numerically optimized tail-recovery bound, and their maximum.
In practice to obtain such a threshold policy with theoretical guarantees, given $\alpha$, choose \(q\) near maximizing the floor objective \((\alpha-q)Q(q)\) or near maximizing \(G_{\mathrm{tail}}(\alpha,q)\) over \(\mathcal D_\alpha\), whichever gives the larger certified bound, and use the threshold \(\tau=Q(q)\). Numerically, the lowest point of the displayed guarantee occurs near the intersection of the two bounds; we do not know whether this crossover reflects a genuine dip in the optimal ratio or only an artifact of combining two lower bounds.

\subsubsection{Proof Outline of Theorem~\ref{thm:draft-prophet-ifra}}

The proof combines two lower bounds on the same class of threshold policies.
 Each lower bound is derived from the same bound \eqref{eq:L_alpha_lower_bound} on \(L_\alpha(S_\tau)\) for a threshold policy \(S_\tau\), but the two bounds use different ways to utilize Assumption~\ref{assump:draft-ifra-prophet} to evaluate the right-hand side of \eqref{eq:L_alpha_lower_bound}.

\emph{Step 1: the threshold-floor estimate.}
We first evaluate the CVaR variational cap at the threshold itself. The elementary threshold calculation in this step uses atomlessness only to identify the no-trigger probability: under Assumption~\ref{assump:draft-ifra-prophet}, \(F_M(Q(q))=q\), so a threshold at \(\tau=Q(q)\) has no-trigger probability \(q\).

Fix \(q\in(0,\alpha)\) and set \(\tau=Q(q)\). For the threshold policy \(S_\tau\), let \(T_\tau\) be the event that the policy accepts an observation, including accepted equality ties. Then \(T_\tau^c\) is the no-trigger event; up to the null event \(\{M=\tau\}\), it is \(\{M\le\tau\}\). Hence \(\pr{T_\tau}=1-F_M(\tau)=1-q\). Moreover, \(S_\tau\ge\tau\) on \(T_\tau\) and \(S_\tau=0\) on \(T_\tau^c\), so \(S_\tau\wedge\tau=\tau\one{T_\tau}\). Using \eqref{eq:cvar-reward} after multiplying by \(\alpha\) and evaluating the cap at \(\tau\),
\[
\begin{aligned}
    L_\alpha(S_\tau)
    &\ge
    \E{S_\tau\wedge\tau}-(1-\alpha)\tau  \\
    &=
    \tau\pr{T_\tau}-(1-\alpha)\tau
    =
    \tau(1-q)-(1-\alpha)\tau
    =
    (\alpha-q)Q(q).
\end{aligned}
\]
The technical consequences of recentered IFRA are stated and proved later in Subsection~\ref{sec:draft-ifra-technical}. In particular, Lemma~\ref{lem:draft-ifra-quantile}(i) shows that \(Q/\ell\) is nonincreasing, and Lemma~\ref{lem:draft-threshold-floor} shows that this monotonicity gives the threshold-floor bound. Applying Lemma~\ref{lem:draft-threshold-floor} with \(h=Q\) and using \(L_\alpha(M)=\int_0^\alpha Q(u)\,du\), we obtain
\begin{equation}
    \sup_\tau
    \frac{\cvar_\alpha(S_\tau)}{\cvar_\alpha(M)}
    =
    \sup_\tau
    \frac{L_\alpha(S_\tau)}{L_\alpha(M)}
    \ge
    \frac{\sup_{0<q<\alpha}(\alpha-q)Q(q)}{L_\alpha(M)}
    \ge
    \rho_{\mathrm{floor}}(\alpha).
    \label{eq:draft-floor-cvar-ratio}
\end{equation}

\emph{Step 2: the tail-recovery estimate.}
Suppose first that \(\mathcal D_\alpha\ne\emptyset\). Fix \(q\in\mathcal D_\alpha\), set \(\tau=Q(q)\), and define \(\gamma_q\defeq 1-(1-\alpha)/q\), so \(1-\alpha=q(1-\gamma_q)\). As in Step 1, the threshold policy with threshold \(\tau\) has no-trigger probability \(q\). Applying \eqref{eq:L_alpha_lower_bound} with \(r=q\) gives, for every \(t\ge\tau\),
\begin{align*}
    L_\alpha(S_\tau)
    &\ge
    \tau(1-q)+q\,\E{(M\wedge t-\tau)_+}-(1-\alpha)t\\
    &\ge
    \tau(1-q)+q\,\E{M\wedge t-\tau}-q(1-\gamma_q)t\\
    &=
    \tau(1-2q)+q\bigl(\E{M\wedge t}-(1-\gamma_q)t\bigr),
\end{align*}
where the second inequality uses \((x)_+\ge x\) and \(1-\alpha=q(1-\gamma_q)\).
We now choose the cap \(t\) in the preceding bound. Because \(q\in\mathcal D_\alpha\), \(q\le\gamma_q\le\alpha\). The lower bound above is valid for every \(t\ge\tau\), and monotonicity of \(Q\) gives \(Q(\gamma_q)\ge Q(q)=\tau\). Thus, if \(\gamma_q<1\), the cap \(t=Q(\gamma_q)\) is feasible and attains the supremum in the variational formula \eqref{eq:cvar-reward} defining \(L_{\gamma_q}(M)\). If \(\gamma_q=1\), the corresponding variational expression is \(\E{M\wedge t}\), and letting \(t\to\infty\) gives \(L_1(M)=\E{M}\).
Hence
\begin{equation}
    L_\alpha(S_\tau)
    \ge
    \tau(1-2q)+qL_{\gamma_q}(M).
    \label{eq:draft-tail-recovery-S-lower}
\end{equation}
Since \(q\in\mathcal D_\alpha\), \(1-2q\ge0\). It turns out in Lemma~\ref{lem:draft-tail-recovery}, stated and proved later in Subsection~\ref{sec:draft-ifra-technical}, that the lower bound \eqref{eq:draft-tail-recovery-S-lower} implies
\[
    \frac{\cvar_\alpha(S_\tau)}{\cvar_\alpha(M)}
    =
    \frac{L_\alpha(S_\tau)}{L_\alpha(M)}
    \ge
    G_{\mathrm{tail}}(\alpha,q).
\]
Taking the supremum over \(q\in\mathcal D_\alpha\),
\[
    \sup_\tau
    \frac{\cvar_\alpha(S_\tau)}{\cvar_\alpha(M)}
    \ge
    \sup_{q\in\mathcal D_\alpha}G_{\mathrm{tail}}(\alpha,q)
    =
    \rho_{\mathrm{tail}}(\alpha),
\]
while if \(\mathcal D_\alpha=\emptyset\), the same inequality is trivial because then \(\rho_{\mathrm{tail}}(\alpha)=0\). Combining this with \eqref{eq:draft-floor-cvar-ratio},
\[
    \sup_\tau
    \frac{\cvar_\alpha(S_\tau)}{\cvar_\alpha(M)}
    \ge
    \max\{\rho_{\mathrm{floor}}(\alpha),\rho_{\mathrm{tail}}(\alpha)\}
    =
    \rho(\alpha).
\]
Therefore, for every \(\varepsilon>0\), some threshold policy attains CVaR ratio at least \(\rho(\alpha)-\varepsilon\).

\subsubsection{Technical Consequences of Recentered IFRA}
\label{sec:draft-ifra-technical}

First, we need an algebraic fact.

\begin{lemma}[Product inequality]
\label{lem:draft-ifra-product}
For \(\lambda\ge1\) and \(p_1,\ldots,p_n\in[0,1]\),
\[
    \prod_{i=1}^n \bigl[1-(1-p_i)^\lambda\bigr]
    \ge
    1-\left(1-\prod_{i=1}^n p_i\right)^\lambda .
\]
\end{lemma}

\begin{myproof}
Let \(f_\lambda(x)=1-(1-x)^\lambda\). The lemma is the statement \(\prod_i f_\lambda(p_i)\ge f_\lambda(\prod_i p_i)\). Thus it suffices to prove the two-variable inequality \(f_\lambda(xy)\le f_\lambda(x)f_\lambda(y)\) for \(x,y\in[0,1]\): applying this inequality repeatedly gives \(f_\lambda(\prod_i p_i)\le f_\lambda(\prod_{i<n}p_i)f_\lambda(p_n)\le\cdots\le\prod_i f_\lambda(p_i)\).

It remains to prove the two-variable inequality. The boundary cases \(x\in\{0,1\}\), \(y\in\{0,1\}\), or \(\lambda=1\) are immediate. Fix \(y\in(0,1)\), and set \(g(x)=f_\lambda(xy)/f_\lambda(x)\). With \(a(z)\defeq z f_\lambda'(z)/f_\lambda(z)=\lambda z(1-z)^{\lambda-1}/[1-(1-z)^\lambda]\), we have \(g'(x)/g(x)=[a(xy)-a(x)]/x\). We claim that \(a\) is nonincreasing. Writing \(s=1-z\), the logarithmic derivative of \((1-s)s^{\lambda-1}/(1-s^\lambda)\) is
\[
    \frac{\lambda-1-\lambda s+s^\lambda}{s(1-s)(1-s^\lambda)}.
\]
Its numerator has derivative \(\lambda(s^{\lambda-1}-1)\le0\) and equals \(0\) at \(s=1\), so it is nonnegative on \((0,1]\). Hence the fraction is nondecreasing in \(s\), and since \(s=1-z\), \(a\) is nonincreasing in \(z\). Therefore \(g'(x)\ge0\). Thus \(f_\lambda(xy)/f_\lambda(x)=g(x)\le g(1)=f_\lambda(y)\), proving the two-variable inequality.
\end{myproof}

\begin{lemma}[IFRA quantile controls]
\label{lem:draft-ifra-quantile}
Under Assumption~\ref{assump:draft-ifra-prophet}, the quantile function \(Q\) of \(M\) satisfies:
\begin{enumerate}
\item[(i)] \(u\mapsto Q(u)/\ell(u)\) is nonincreasing on \((0,1)\).
\item[(ii)] For \(0<q<\alpha\le1\),
\[
    L_\alpha(M)
    \le
    Q(q)\frac{\Psi(\alpha)+\ell(q)-q}{\ell(q)}.
\]
\item[(iii)] For \(0<\gamma\le\alpha\le1\),
\[
    \frac{L_\gamma(M)}{L_\alpha(M)}
    \ge
    \frac{\Psi(\gamma)}{\Psi(\alpha)}.
\]
\end{enumerate}
\end{lemma}

\begin{myproof}
For (i), let \(F_i\) and \(H_i\) denote the cumulative distribution function and cumulative hazard of \(X_i\), respectively. We claim that Definition~\ref{def:draft-ifra} implies that, for \(\lambda\ge1\),
\[
    F_i(\lambda x)\ge 1-(1-F_i(x))^\lambda .
\]
Indeed, the boundary cases \(F_i(x)\in\{0,1\}\) or \(\lambda x\ge u_i\) are immediate. Otherwise \(x\in(s_i,u_i)\), \(\lambda x<u_i\), and
\[
    H_i(\lambda x)
    \ge
    \frac{\lambda x-s_i}{x-s_i}H_i(x)
    \ge
    \lambda H_i(x),
\]
where the last inequality uses \(s_i\ge0\). Exponentiating gives the scaling inequality. Hence, using independence and Lemma~\ref{lem:draft-ifra-product} with \(p_i=F_i(x)\),
\begin{equation}
    F_M(\lambda x)
    =
    \prod_{i=1}^n F_i(\lambda x)
    \ge
    \prod_{i=1}^n\brackets{1-(1-F_i(x))^\lambda}
    \ge
    1-\paren{1-\prod_{i=1}^nF_i(x)}^\lambda
    =
    1-(1-F_M(x))^\lambda .
    \label{eq:draft-ifra-max-scaling}
\end{equation}
For \(0<q\le v<1\), set \(x=Q(q)\) and \(\lambda=\ell(v)/\ell(q)\). Since \(\ell\) is increasing, \(\lambda\ge1\), and \(F_M(Q(q))=q\), \eqref{eq:draft-ifra-max-scaling} gives
\[
    F_M\paren{\frac{\ell(v)}{\ell(q)}Q(q)}
    \ge
    1-\paren{1-q}^{\ell(v)/\ell(q)}
    =
    1-\exp\{-\ell(v)\}
    =
    v .
\]
By the definition of the quantile function, \(Q(v)\le Q(q)\ell(v)/\ell(q)\), or equivalently \(Q(v)/\ell(v)\le Q(q)/\ell(q)\). This proves (i).

For (ii), monotonicity of \(Q\) gives \(Q(u)\le Q(q)\) for \(u\le q\), while part (i) gives \(Q(u)\le Q(q)\ell(u)/\ell(q)\) for \(u\ge q\). Hence, using \(\int_q^\alpha\ell(u)\,du=\Psi(\alpha)-\Psi(q)\),
\begin{align*}
    L_\alpha(M)
    &=
    \int_0^\alpha Q(u)\,du
    \le
    qQ(q)+\frac{Q(q)}{\ell(q)}\int_q^\alpha \ell(u)\,du\\
    &=
    \frac{Q(q)}{\ell(q)}\paren{\Psi(\alpha)+q\ell(q)-\Psi(q)}
    =
    Q(q)\frac{\Psi(\alpha)+\ell(q)-q}{\ell(q)},
\end{align*}
where the last equality uses \(q\ell(q)-\Psi(q)=\ell(q)-q\).

For (iii), write \(Q(u)=r(u)\ell(u)\), where \(r\) is nonincreasing by part (i). Then
\[
    A(z)\defeq\frac{L_z(M)}{\Psi(z)}
    =
    \frac{\int_0^z r(u)\ell(u)\,du}{\int_0^z \ell(u)\,du}
\]
is a weighted average of \(r\) over \((0,z)\). For \(0<a<b\), let
\(A_{a,b}\defeq \int_a^b r(u)\ell(u)\,du/\int_a^b \ell(u)\,du\). Since \(r\) is nonincreasing and \(\ell>0\), \(A_{a,b}\le r(a)\le A(a)\). Hence
\[
    A(b)
    =
    \frac{\Psi(a)}{\Psi(b)}A(a)
    +
    \frac{\Psi(b)-\Psi(a)}{\Psi(b)}A_{a,b}
    \le
    A(a),
\]
so \(A\) is nonincreasing. Since \(0<\gamma\le\alpha\), this gives \(A(\gamma)\ge A(\alpha)\), proving (iii).
\end{myproof}

\begin{lemma}[Threshold-floor bound]
\label{lem:draft-threshold-floor}
Let \(\alpha\in(0,1]\), and let \(h:(0,\alpha)\to[0,\infty)\) be a deterministic function. Suppose that \(h\) is nondecreasing, that \(h/\ell\) is nonincreasing on \((0,\alpha)\), and that \(\int_0^\alpha h(u)\,du<\infty\). Then
\[
    \sup_{0<q<\alpha}(\alpha-q)h(q)
    \ge
    \rho_{\mathrm{floor}}(\alpha)\int_0^\alpha h(u)\,du.
\]
\end{lemma}

\begin{myproof}
If \(\int_0^\alpha h(u)\,du=0\), then \(h\equiv0\) on \((0,\alpha)\), and the claim is immediate. Otherwise rescale \(h\) so that \(\int_0^\alpha h(u)\,du=1\). By the definition of the supremum,
\begin{equation}
    h(q)
    \le
    \frac{\sup_{0<p<\alpha}(\alpha-p)h(p)}{\alpha-q},
    \qquad 0<q<\alpha.
    \label{eq:draft-floor-sup-bound}
\end{equation}
The quantity \(q_\alpha\) is the unique maximizer of \(f(q)=(\alpha-q)\ell(q)\) on \((0,\alpha)\): indeed,
\[
    f'(q)=-\ell(q)+\frac{\alpha-q}{1-q},
    \qquad
    f''(q)=\frac{\alpha+q-2}{(1-q)^2}<0.
\]
For \(q\ge q_\alpha\), monotonicity of \(h/\ell\) and \eqref{eq:draft-floor-sup-bound} at \(q_\alpha\) give
\begin{equation}
    \frac{h(q)}{\ell(q)}
    \le
    \frac{h(q_\alpha)}{\ell(q_\alpha)}
    \le
    \frac{\sup_{0<p<\alpha}(\alpha-p)h(p)}
    {(\alpha-q_\alpha)\ell(q_\alpha)}.
    \label{eq:draft-floor-tail-bound}
\end{equation}
Integrating \(h\) with \eqref{eq:draft-floor-sup-bound} on \((0,q_\alpha)\) and \eqref{eq:draft-floor-tail-bound} on \((q_\alpha,\alpha)\),
\begin{align*}
    1
    =
    \int_0^\alpha h(u)\,du
    &\le
    \sup_{0<p<\alpha}(\alpha-p)h(p)
    \int_0^{q_\alpha}\frac{du}{\alpha-u}
    +
    \frac{\sup_{0<p<\alpha}(\alpha-p)h(p)}
    {(\alpha-q_\alpha)\ell(q_\alpha)}
    \int_{q_\alpha}^{\alpha}\ell(u)\,du\\
    &=
    \sup_{0<p<\alpha}(\alpha-p)h(p)
    \left[
        \log\frac{\alpha}{\alpha-q_\alpha}
        +
        \frac{\Psi(\alpha)-\Psi(q_\alpha)}
        {(\alpha-q_\alpha)\ell(q_\alpha)}
    \right]
    =
    \sup_{0<p<\alpha}(\alpha-p)h(p)\,
    \rho_{\mathrm{floor}}(\alpha)^{-1}.
\end{align*}
Thus \(\sup_{0<p<\alpha}(\alpha-p)h(p)\ge\rho_{\mathrm{floor}}(\alpha)\) under the normalization, and scaling back proves the result.
\end{myproof}

\begin{lemma}[Tail-recovery IFRA comparison]
\label{lem:draft-tail-recovery}
Under Assumption~\ref{assump:draft-ifra-prophet}, fix \(q\in\mathcal D_\alpha\) and set \(\tau=Q(q)\). If a threshold payoff \(S_\tau\) satisfies
\begin{equation}
    L_\alpha(S_\tau)
    \ge
    \tau(1-2q)+qL_{1-(1-\alpha)/q}(M),
    \label{eq:draft-tail-recovery-hypothesis}
\end{equation}
then
\[
    \frac{\cvar_\alpha(S_\tau)}{\cvar_\alpha(M)}
    =
    \frac{L_\alpha(S_\tau)}{L_\alpha(M)}
    \ge
    G_{\mathrm{tail}}(\alpha,q).
\]
\end{lemma}

\begin{myproof}
Since \(q\in\mathcal D_\alpha\), \(1-2q\ge0\). Lemma~\ref{lem:draft-ifra-quantile}(ii) gives
\[
    \frac{\tau}{L_\alpha(M)}
    \ge
    \frac{\ell(q)}{\Psi(\alpha)+\ell(q)-q},
\]
The level \(1-(1-\alpha)/q\) is feasible because \(q\in\mathcal D_\alpha\): indeed,
\[
    q(1-q)\ge 1-\alpha
    \quad\Longrightarrow\quad
    1-\frac{1-\alpha}{q}\ge q>0,
\]
and, since \(q<\alpha\le1\), \((1-\alpha)/q\ge1-\alpha\), hence \(1-(1-\alpha)/q\le\alpha\) (with equality when \(\alpha=1\)). Thus Lemma~\ref{lem:draft-ifra-quantile}(iii) gives
\[
    \frac{L_{1-(1-\alpha)/q}(M)}{L_\alpha(M)}
    \ge
    \frac{\Psi(1-(1-\alpha)/q)}{\Psi(\alpha)}.
\]
Substituting these two estimates into \eqref{eq:draft-tail-recovery-hypothesis} and dividing by \(L_\alpha(M)\) yields
\[
    \frac{\cvar_\alpha(S_\tau)}{\cvar_\alpha(M)}
    =
    \frac{L_\alpha(S_\tau)}{L_\alpha(M)}
    \ge
    (1-2q)\frac{\ell(q)}{\Psi(\alpha)+\ell(q)-q}
    +
    q\,\frac{\Psi(1-(1-\alpha)/q)}{\Psi(\alpha)}
    =
    G_{\mathrm{tail}}(\alpha,q).
\]
\end{myproof}

\bibliographystyle{plainnat}
\bibliography{ref}

\end{document}